\documentclass[aps,pra,twocolumn,superscriptaddress,showpacs,showkeys,amsmath,amssymb]{revtex4}
\usepackage{amsfonts}
\usepackage{mathrsfs}
\usepackage{latexsym}
\usepackage[cp1251]{inputenc}
\usepackage{graphicx}
\usepackage{dcolumn}
\usepackage{bm}
\usepackage{color}
\RequirePackage{ifthen}
\RequirePackage[pdfstartview=FitH]{hyperref}

\begin{document}

    \title{Large-$N$ properties of a non-ideal Bose gas}
    \author{Orest~Hryhorchak}
    \author{Volodymyr~Pastukhov\footnote{e-mail: volodyapastukhov@gmail.com}}
    \affiliation{Department for Theoretical Physics, Ivan Franko National University of Lviv,\\ 12 Drahomanov Street, Lviv-5, 79005, Ukraine}

    \date{\today}



    \begin{abstract}
We rigorously discuss the large-$N$ thermodynamics of a Bose gas with a short-range two-body potential. Considering the system as a mixture of $N$ identical components with symmetrical interaction we calculated numerically the temperature dependence of the leading-order corrections to the depletion of Bose-Einstein condensate and to the isothermal compressibility.
    \end{abstract}

    \maketitle

\section{Introduction}
\label{sec1} \setcounter{equation}{0}
The $1/N$ expansion is known to be a powerful tool in the field theory \cite{Moshe} and in the condensed matter physics, where it allowed to describe the properties of frustrated quantum antiferromagnets \cite{Read}, strongly-coupled low-density quantum gases \cite{Nikolic}, unitary Fermi superfluids \cite{Veillette}, fermions possessing the finite-temperature Stoner instability \cite{He}, graphene \cite{Son,Foster}, impurity states in the dilute critical Bose condensate \cite{Pastukhov}, etc. It has also been proven as a valuable method for the approximate calculations of critical exponents \cite{Vasilev}. In the large-$N$ limit some of field-theoretical \cite{Brezin} and statistic-mechanical \cite{Stanley} models can be solved exactly and every time the behavior of these systems is highly non-trivial \cite{Brezin_Wadia}. 

The large-$N$ expansion in the context of Bose systems' theory was previously used in different variations including summation of diagrams in a particle-hole channel for the Galilean-invariant \cite{Andersen,Nogueira}, relativistic \cite{Brauner}, and two-component \cite{Chien} many-boson systems as well as ladder summation of particle-particle diagrams for the normal state Bose gases \cite{Liu}, but the full numerical analysis of the problem at finite temperatures in the Bose condensed phase is lacking. The application of this method to two-dimensional Bose systems \cite{Nogueira_Kleinert} revealed very interesting phenomenon, namely, the appearance of thermally-stimulated roton minimum in the spectrum of collective excitations. Recently, we have demonstrated \cite{Hryhorchak} that the application of this technique to the problem of critical temperature calculation for a Bose gas with point-like repulsive interaction leads to the result which coincides semi-quantitatively well with that of the Monte Carlo simulations \cite{Bronin,Nguyen} in a whole range of the interaction parameter. The latter observation inspires confidence in the use of the large-$N$-based expansion for studying of the finite-temperature thermodynamic properties of Bose gas not only in the dilute limit.

\section{Model and method}
We consider Bose mixture consisting of $N$ identical components with symmetric zero-range interparticle interaction $\Phi(r)=(g/N)\delta({\bf r})\frac{\partial}{\partial r}r$, where the coupling constant is related to the $s$-wave scattering length $a$ and mass of particles $m$ in the following way: $g=4\pi\hbar^2a/m$. This model after introducing auxiliary real field $\varphi(x)$ and making use of the Hubbard-Stratonovich transformation is described by the following imaginary time action \cite{Hryhorchak}
\begin{eqnarray}\label{S}
	S=\int dx\, \psi^*_{\sigma}(x)\left\{\partial_{\tau}-\xi-i\varphi(x)\right\}
	\psi_{\sigma}(x)\nonumber\\
	-\frac{N}{2g}\int dx
	\varphi^2(x),
\end{eqnarray}
where $x\equiv(\tau, {\bf r})$, $\xi=-\hbar^2\nabla^2/2m-\mu$, $\int
dx=\int_0^{1/T}d\tau\int_Vd{\bf r}$, periodic complex fields $\psi_{\sigma}(x)$ describe $N$ sorts of Bose particles those number is controlled by chemical potential $\mu$ and the summation over repeating index $\sigma=1,2,\ldots ,N$ is
implied. As usual, in the symmetry broken phase one singles out the condensates of two fields $\psi_{\sigma}(x)=\psi_0+\tilde{\psi}_{\sigma}(x)$, 
$\varphi(x)=\varphi_0+\tilde{\varphi}(x)$ with constraints $\int_V
d{\bf r}\,\tilde{\psi}_{\sigma}(x)=\int_V
d{\bf r}\,\tilde{\varphi}(x)=0$ imposed on the fluctuational terms. The steepest descend
method for the grand canonical potential $\left(\partial \Omega/\partial
\varphi_0\right)_{\psi_0}=0$, $\left(\partial \Omega/\partial
\psi^*_0\right)_{\varphi_0}=0$ fixes the value of $i\varphi_0=ng$ ($n$ is the density of each sort of particles) and generates the ``gap'' equation
\begin{eqnarray}\label{gap_eq}
\mu\psi_0-ng\psi_0-\frac{iT}{V}\int dx\,\langle\tilde{\varphi}(x)\tilde{\psi}_{\sigma}(x)\rangle=0,
\end{eqnarray}
where $T$ is the temperature and $\langle\ldots \rangle$ denotes statistical averaging. The above equation possesses at least two physically distinct solutions. The trivial one $\psi_0=0$ describes properties of the system in the high-temperature region while the second solution determines the chemical potential in the Bose condensed phase.

Action (\ref{S}) after the shift of fields $\varphi(x)$, $\psi_{\sigma}(x)$  in the four-momentum space reads
\begin{align}\label{S_F}
& S=\sum_P\left\{i\omega_p-\tilde{\xi}_p \right\}\psi^{*}_{\sigma,P}\psi_{\sigma,P}-\frac{N}{2g}\sum_Q|\varphi_Q|^2\nonumber\\
&-i\sum_{Q,\sigma}\varphi_Q\{\psi^*_0\psi_{\sigma,-Q}+\psi^*_{\sigma,Q}\psi_0\}\nonumber\\
&-i\sqrt{\frac{T}{V}}\sum_{Q,P}\varphi_Q\psi^{*}_{\sigma,P}\psi_{\sigma,P-Q},
\end{align}
where capital letters denote four-momenta $P\equiv(\omega_p, {\bf
p}\neq 0)$,  $Q\equiv(\omega_q, {\bf q}\neq 0)$ (here
$\omega_q,\omega_p$ are bosonic Matsubara frequencies) and $\tilde{\xi}_p=\varepsilon_p-\tilde{\mu}$ with
$\varepsilon_p=\hbar^2p^2/2m$ being the free-particle dispersion. We also
used notation for the shifted chemical potential
$\tilde{\mu}=\mu-ng$.

The action (\ref{S_F}) without the last term is a quadratic form over fields $\psi_{\sigma, P}$, $\varphi_Q$ and the problem of calculation of the appropriate averages $\langle\psi^*_{\sigma,P} \psi_{\sigma',P}\rangle$, $\langle\psi_{\sigma,P} \psi_{\sigma,-P}\rangle$, $\langle\psi^*_{\sigma,P} \varphi_{P}\rangle$, etc. reduces to the diagonalization of quadratic matrix of size $2N+1$. It is important to note that these calculations can be exactly performed in general case for an arbitrary $N$ with the result
\begin{eqnarray}\label{psi_psi}
\langle\psi^*_{\sigma,P} \psi_{\sigma',P}\rangle=\frac{\delta_{\sigma, \sigma'}}{\tilde{\xi}_p-i\omega_p}-\frac{1}{N}\frac{\tilde{\xi}_p+i\omega_p}{\tilde{\xi}_p-i\omega_p}\frac{|\psi_0|^2g}{\omega^2_p+E^2_p},
\end{eqnarray}
\begin{eqnarray}\label{psi_psi_a}
\langle\psi_{\sigma,P} \psi_{\sigma',-P}\rangle=-\frac{1}{N}\frac{\psi^2_0g}{\omega^2_p+E^2_p},
\end{eqnarray}
\begin{eqnarray}\label{psi_varphi}
\langle\psi_{\sigma,P} \varphi_{-P}\rangle=-\frac{i\psi_0g}{N}\frac{\tilde{\xi}_p-i\omega_p}{\omega^2_p+E^2_p},
\end{eqnarray}
\begin{eqnarray}
\langle\varphi_{Q}\label{varphi_varphi} \varphi_{-Q}\rangle=\frac{g}{N}\frac{\omega^2_q+\tilde{\xi}^2_q}{\omega^2_q+E^2_q},
\end{eqnarray}
where $\delta_{\sigma, \sigma'}$ is the Kronecker delta. At this stage the formulated approximation is nothing but the celebrated Bogoliubov theory for $N$-component Bose system with symmetric inter- and intra-species interactions and characteristic quasiparticle spectrum $E^2_q=\tilde{\xi}^2_q+2\tilde{\xi}_qn_0g$ (where $n_0\equiv|\psi_0|^2$). Furthermore, by setting $N=1$ in Eqs.~(\ref{psi_psi}), (\ref{psi_psi_a}) we recover (up to a sign, of course) the well-known expressions (see, for instance \cite{Popov}) for the normal and anomalous Green's functions in the simplest approximation. The inclusion of last term in action brakes the exact integrability of the partition function, but in the leading order over the expansion parameter $1/N$ the result is easily obtained. To do this, one has to replace in above formulas the bare interaction parameter $g$ by the density-fluctuation-induced effective potential $g(Q)\equiv g/[1+g\Pi(Q)]$, 
\begin{figure}[h!]
	\centerline{\includegraphics
		[width=0.47\textwidth,clip,angle=-0]{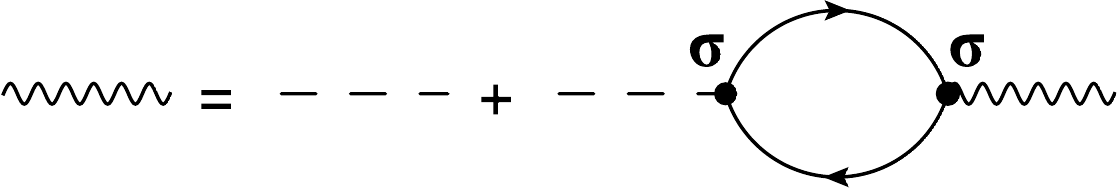}}
	\caption{The diagrammatic equation that determines effective potential $g(K)$ (wavy line) to the leading order. The dashed line stands for $g$, dots are bare vertices equal to $-i$, and arrows represent the ideal Bose gas correlators $\langle\psi^*_{\sigma,P} \psi_{\sigma,P}\rangle$ (the first term in Eq.~(\ref{psi_psi})).}
\end{figure}
where the polarization operator reads
\begin{eqnarray}\label{Pi}
\Pi(Q)=\frac{T}{V}\sum_P\frac{1}{\tilde{\xi}_p-i\omega_p}\frac{1}{\tilde{\xi}_{|{\bf p}+{\bf q}|}-i\omega_{p+q}}.
\end{eqnarray}

Additionally, we have to shift $\tilde{\xi}_p\to \tilde{\xi}_p+\Sigma^{(1)}(P)/N$ the denominator in the first term of Eq.~(\ref{psi_psi}) by the $1/N$-correction to the diagonal element of self-energy (see Fig.~2) $\Sigma^{(1)}(P)$, which in our approximation equals to
\begin{eqnarray}\label{Sigma}
\Sigma^{(1)}(P)=\frac{T}{V}\sum_Q \frac{g(Q)(\tilde{\xi}^2_q+\omega^2_q)}{E^2(Q)+\omega^2_q}\frac{1}{\tilde{\xi}_{|{\bf p}+{\bf q}|}-i\omega_{p+q}},
\end{eqnarray}
\begin{figure}[h!]
	\centerline{\includegraphics
		[width=0.23\textwidth,clip,angle=-0]{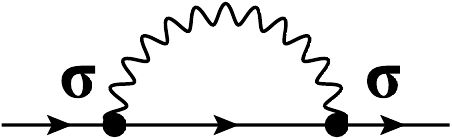}}
	\caption{The $1/N$ self-energy $\Sigma^{(1)}(P)$ (notations are similar to that in Fig.~1). The order of diagram in the large-$N$ expansion is determined by the number of wavy lines. Note that diagrams with different vertex indices $\sigma$ (non-diagonal elements of self-energy) are of order $1/N^2$, because correlator $\langle\psi^*_{\sigma,P} \psi_{\sigma',P}\rangle$ with $\sigma \neq \sigma'$ itself contains factor $1/N$ (see Eq.~(\ref{psi_psi})). The same concerns the matrix of anomalous self-energies, where all elements are of order $1/N^2$.}
\end{figure}
(where $E(Q)=E_q|_{g\to g(Q)}$). A very important feature of our approximation is that it preserves the Nepomnyashchy-Nepomnyashchy identity \cite{Nepomnyashchy} for the self-energy of normal Green's function. Indeed, taking the limit $P\to 0$ in the above self-energy and substituting Eq.~(\ref{psi_varphi}) (with the replacement $g\to g(Q)$) in the ``gap'' equation (\ref{gap_eq}) one readily obtains that $\tilde{\mu}=\Sigma^{(1)}(0)/N$. This guarantees for the one-particle spectrum (which coincides with the spectrum of collective modes in this approximation), in accordance to the Hugengoltz-Pines relation \cite{Hugenholtz_Pines}, to be gapless in the Bose condensate phase. At high temperatures the quantity $\Sigma^{(1)}(0)/N-\tilde{\mu}$ is always positive definite and up to the leading order in the expansion over $1/N$ plays a role of an effective chemical potential that signals the occurrence of the Bose-Einstein condensation (BEC) transition. A very similar parameter appears in extension of the Beliaev technique \cite{Capogrosso-Sansone} on the finite-temperature region. Finally it should be emphasized that the formulated approach has much in common with the so-called dielectric formalism \cite{Szepfalusy,Fliesser_etal,Navez} applied to Bose condensed systems, where the analogue of our leading-order $1/N$-treatment is the celebrated Random Phase Approximation (RPA). But despite dielectric formalism the large-$N$ technique allows to perform controllable approximate calculations in terms of formal dimensionless expansion parameter $1/N$, which leads to the disparity in results of two approaches even in the simplest approximation. Particularly a very important self-energy correction (\ref{Sigma}), which is of order $1/N$ and renormalizes the one-particle spectrum as well as shifts the critical temperature of an interacting Bose system is not taken into account in the original RPA. A possible resolution of this problem can be found in various generalizations of the RPA, which were comprehensively studied in Refs.~\cite{Watabe}.

\section{Results and discussion}
Having calculated correlator (\ref{psi_psi}) up to $1/N$-terms we are in position to obtain the thermodynamic functions and the Bose condensate depletion. The main differences between large-$N$ expansion and Bogoliubov's theory are expected at finite temperatures, especially in the narrow region of critical point, where developed thermal fluctuations have a profound effect on the behavior of Bose systems.

Both the temperature and interaction-induced depletion of the BEC density of each sort can be calculated in a standard manner
\begin{eqnarray}\label{n_0}
n_0=n-\lim_{\tau \to +0}\frac{T}{V}\sum_{P}e^{i\tau\omega_p}\langle\psi^*_{\sigma,P} \psi_{\sigma,P}\rangle,
\end{eqnarray}
where in the adopted approximation we have to substitute $\langle\psi^*_{\sigma,P} \psi_{\sigma,P}\rangle=(2.4)-\frac{1}{N}\Sigma^{(1)}(P)/(\tilde{\xi}_p-i\omega_p)^2$ (recall $g\to g(Q)$) and to replace $\tilde{\mu}\to 0$ in all $1/N$-terms. We also had to take into account the shift of critical temperature up to order $1/N$. The numerical computations of the temperature dependence of $n_0$ (see Fig.~3) were performed at $N=1$ and for three values of gas parameter, namely $an^{1/3}=0.01$, $0.125$ and $0.345$.
\begin{figure}[h!]
	\centerline{\includegraphics
		[width=0.6\textwidth,clip,angle=-0]{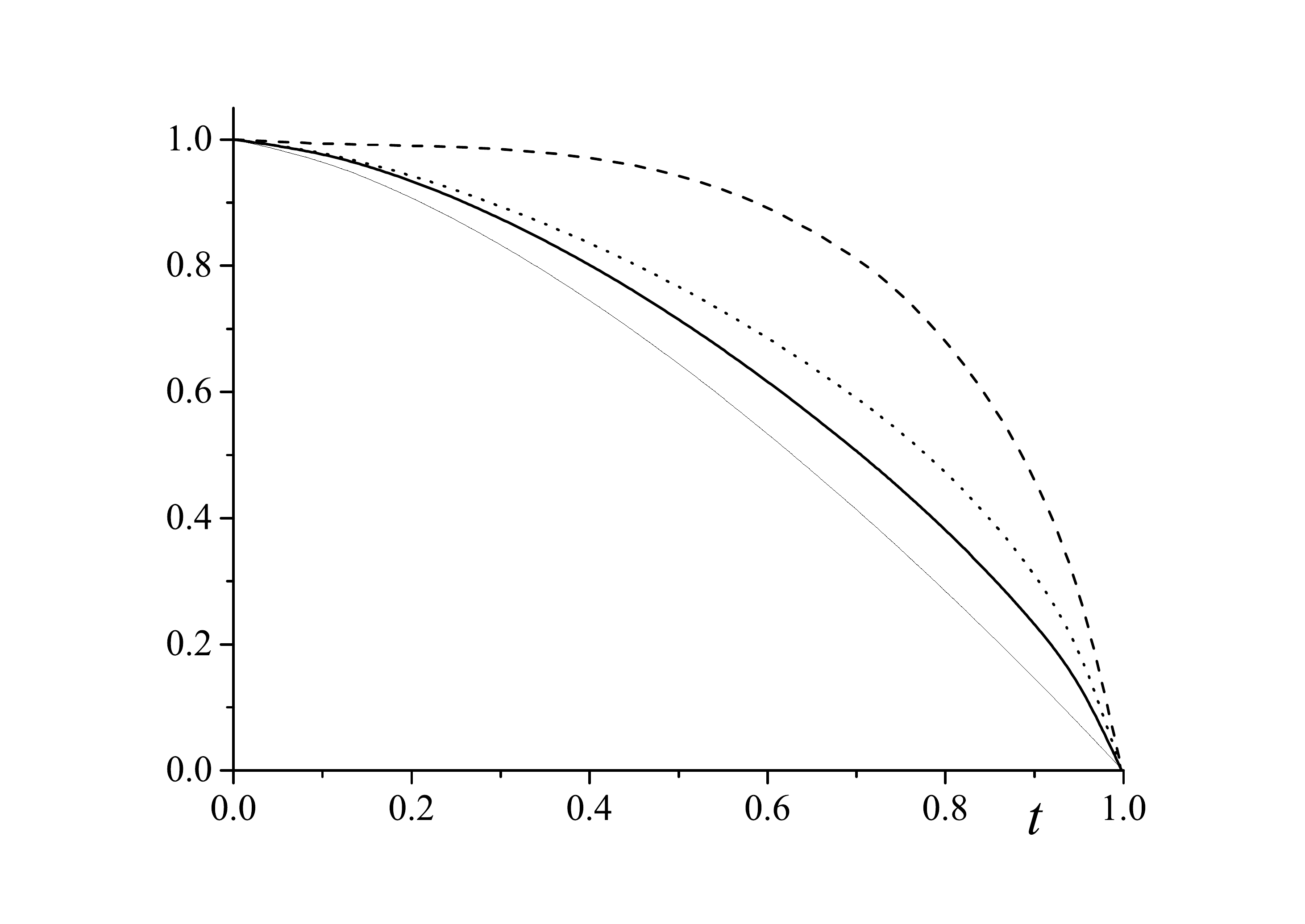}}
	\caption{The temperature dependence (in dimensionless units $t=T/T_c$) of the Bose condensate rescaled on zero-temperature Bogoliubov's result  $n_0=n[1-8\sqrt{na^3}/(3\sqrt{\pi})]$. Solid, dotted and dashed lines correspond to values of gas parameter $an^{1/3}=0.01$, $an^{1/3}=0.125$ and $an^{1/3}=0.345$, respectively. The lower thin line depicts the condensate fraction for ideal Bose gas.}
\end{figure}
The last two correspond to maximum of the critical temperature shift and to the value where it becomes zero again \cite{Hryhorchak}, respectively. From the obtained graphical dependences of condensate density the general prosperity is clearly viewed: the increase of two-body interaction lifts up the condensate curve rescaled on the ground-state (Bogoliubov's in our case) value above the ideal gas result. This conclusion of the large-$N$ approach should be compared with the various finite-temperature self-consistent beyond-Bogoliubov treatments \cite{Kita,Yukalovs_1,Cooper_etal,Zhang,Yukalovs_2} and numerical approaches \cite{Brewczyk_etal}. Moreover, recent experimental test of the Bogoliubov ground-state depletion \cite{Lopes_etal} gives a hope for the forthcoming verification of our finite-temperature results. The distinct feature of the $1/N$ expansion is that it possesses the non-trivial critical behavior. In particular, it is easy to show by analyzing the zero Matsuraba frequency term in Eq.~(\ref{n_0}) that the condensate depletion demonstrates a logarithmic non-analyticity in the close vicinity of BEC transition temperature $T_c$
\begin{eqnarray}\label{n_0_cr}
n_0/n\propto \delta t-\frac{4}{N\pi^2} \delta t\ln \delta t+\ldots, 
\end{eqnarray}
(here $\delta t=\frac{T_c-T}{T_c}$) where the universal power-law behavior of the order parameter $\psi_0\sim (\delta t)^{\beta}$ is clearly visible. Of course, the calculated here exponent $\beta=1/2-2/(N\pi^2)\mathop{\longrightarrow}\limits_{N=1} 0.2973\ldots$ reproduces the large-$N$ result of Ref.~\cite{Abe}, but is far from the values $0.3485(2)$ \cite{Campostrini_01} and $0.3486(1)$ \cite{Campostrini_06} obtained for a single-component Bose system's universality class in Monte Carlo simulations. These findings, however, together with the value of the Fisher exponent $\eta=4/(3N\pi^2)$ \cite{Vakarchuk_etal,Pastukhov} obtained with the same accuracy determine the critical behavior of the three-dimensional Bose system in the large-$N$ limit.

In the present article we also focus on the derivation of the $1/N$-correction to inverse susceptibility $(\partial \mu/\partial n)_T$. This quantity determines the velocity of the first sound $c^2=\frac{n}{m}(\partial \mu/\partial n)_T$ in the Bose condensate and can be measured in the appropriate experiments. The simplest way to deal with the problem of its calculation in the Bose-condensed phase is first to obtain the beyond-mean-field  correction to the chemical potential with the use of Eq.~(\ref{gap_eq}) (or equivalently Eq.~(\ref{Sigma}))
\begin{eqnarray}
\mu=ng+\mu^{(1)}/N,
\end{eqnarray}
where $1/N$-shift is given by
\begin{eqnarray}\label{mu_1}
\mu^{(1)}=\frac{T}{V}\sum_Q \left(\frac{g(Q)\varepsilon_q}{E^2(Q)+\omega^2_q}-\frac{g\varepsilon_q}{E^2_q+\omega^2_q}\right)\nonumber\\
+\frac{1}{2V}\sum_{{\bf q}\neq 0}g\left(\frac{\varepsilon_q}{E_q}-1\right)
+\frac{1}{V}\sum_{{\bf q}\neq 0}\frac{g\varepsilon_q}{E_q}n(E_q/T),
\end{eqnarray}
here $n(x)=1/(e^x-1)$ is the Bose distribution and we singled out standard Bogoliubov's corrections (last two terms) explicitly, and then to evaluate straightforwardly the derivative with respect to $n$. 

The results of numerical calculations for the temperature dependence of the compressibility are presented in Fig.~4.
\begin{figure}[h!]
    \centerline{\includegraphics
        [width=0.6\textwidth,clip,angle=-0]{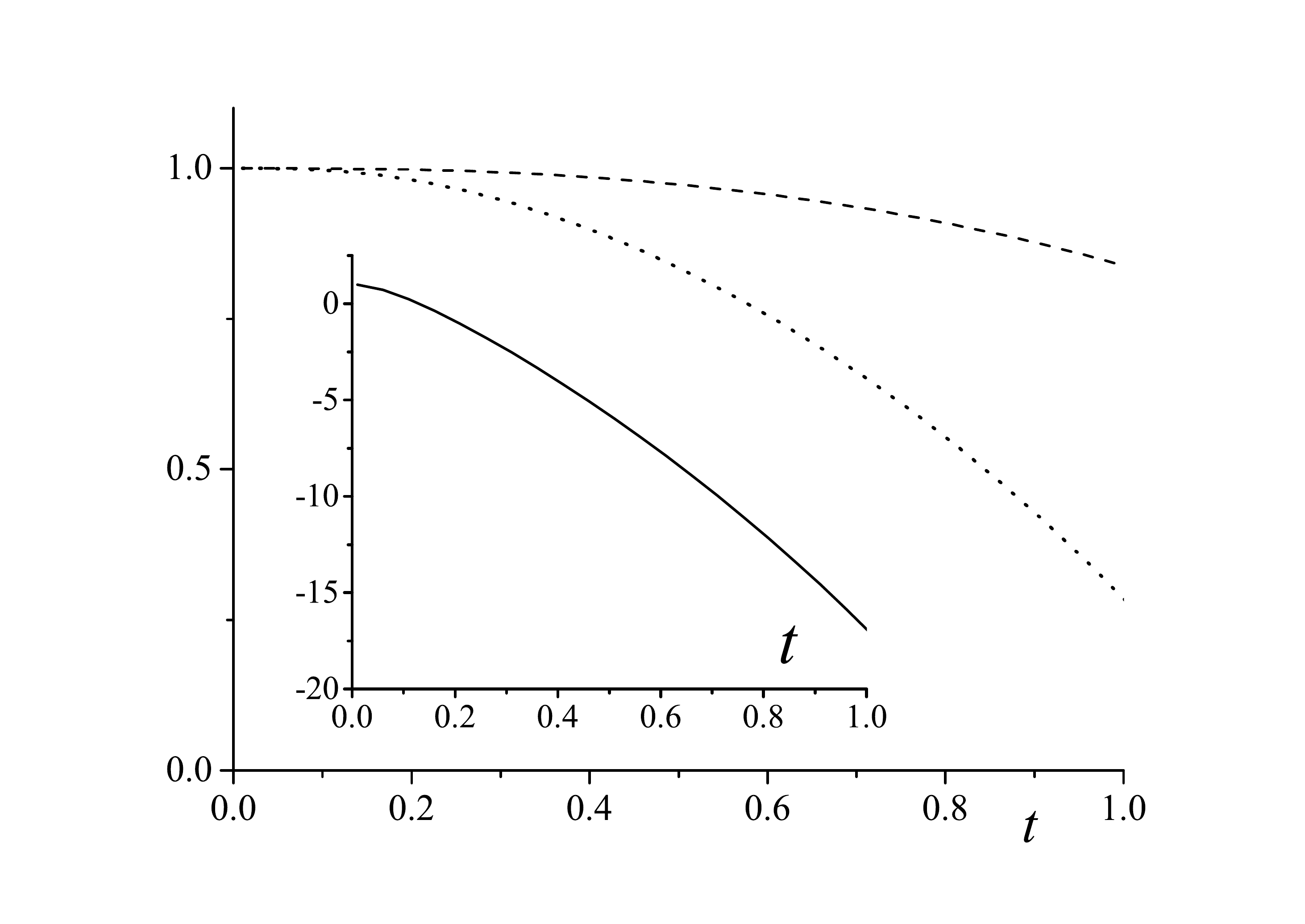}}
   \caption{The temperature ($t=T/T_c$) dependence of the $1/N$ correction to compressibility $(\partial\mu^{(1)}/\partial n)_T$ of Bose gas in the condensed phase rescaled on the zero-temperature (Lee-Huang-Yang) result $16g\sqrt{na^3}$. Dotted and dashed lines correspond to values of the gas parameter $an^{1/3}=0.125$ and $an^{1/3}=0.345$, respectively. The inset shows result for the dilute limit $an^{1/3}=0.01$.}
\end{figure}
In order to compare regimes of various coupling strength we have chosen the same values of the gas parameter as previously used for the condensate calculations. From general arguments it is become clear that the temperature effects should be more visible in the behavior of the sound velocity of a dilute Bose gas, while in the case of intermediate interaction strength the compressibility due to increasing role of phonon excitations is less sensitive to the temperature changes. These intuitive conclusions are confirmed by direct numerical computations and the general tendency of obtained curves is in the qualitative agreement with the observed temperature behavior of the first sound velocity in liquid $^4$He \cite{Hryhorchak_15}. As it is seen from Fig.~4 the calculated isothermal compressibility is finite at $T_c$, but one can easily show that likewise the condensate depletion (\ref{n_0_cr}), its leading-order temperature dependence is log-linear in the vicinity of the critical point. Obviously this is a hint for an universal power-law behavior of the inverse susceptibility close to the BEC transition temperature.

\section{Conclusions}
In summary, by means of the $1/N$-expansion we studied the thermodynamic properties of Bose gas in the condensed phase. At absolute zero this approach in the simplest approximation recovers the well-known results for thermodynamics of a weakly-interacting Bose gas, but the presence of a formal small parameter, namely, the inverse number of components allows to perform controllable calculations not only in the dilute limit. The main differences of our consideration in comparison with the conventional Bogoliubov theory were found in the finite-temperature region, where the large-$N$ expansion provides the qualitatively correct shift of the Bose-Einstein condensation critical temperature, changes an order of phase transition and generally impacts crucially on thermodynamics of the system. Particularly working in the first order over the expansion parameter we determined the full temperature dependence of the condensate density for a Bose system with the point-like repulsion. By varying interparticle interaction we were able to cover both the dilute limit and the case of intermediate couplings for which the calculated in present article temperature behavior of the Bose condensate is in qualitative agreement with the observed (see, for instance \cite{Boninsegni}) in Monte Carlo simulations dependence of condensate fraction in the superfluid helium. The evaluation of inverse susceptibility revealed a natural tendency of sound velocity to become less sensitive to the thermal effects with increasing strength of the repulsive two-body potential. The proposed approach, after some modifications, can be used for the describing of recent experiment \cite{Ville} which has made available the measurements of the finite-temperature behavior of sound velocity in the uniform two-dimensional Bose gas. Finally, it is meanwhile shown that the large-$N$ expansion is capable to capture the correct temperature dependence of basic thermodynamic functions of Bose condensates both in the low-temperature limit and in narrow region of the critical point, and can be used for describing future experiments with uniform quantum gases.

\begin{center}
{\bf Acknowledgements}
\end{center}
The authors thank to Profs. Ivan Vakarchuk and Andrij Rovenchak for invaluable comments. This work was partly supported by Project FF-30F (No.~0116U001539)
from the Ministry of Education and Science of Ukraine.

%\section{Appendices}
%\subsection{Self-energy}


\begin{thebibliography}{99}
	
	
\bibitem{Moshe} M.~Moshe and J.~Zinn-Justin, Phys.~Rep. {\bf 385}, 69 (2003).
\bibitem{Read} N.~Read and S.~Sachdev, Phys.~Rev.~Lett. {\bf 66}, 1773  (1991).
\bibitem{Nikolic} P.~Nikoli\'{c} and S.~Sachdev, Phys.~Rev.~A {\bf 75}, 033608 (2007).
\bibitem{Veillette} M.~Y.~Veillette, D.~E.~Sheehy, and L.~Radzihovsky, Phys.~Rev.~A {\bf 75}, 043614 (2007).
\bibitem{He} L.~He, X.-J. Liu, X.-G. Huang, and H.~Hu, Phys.~Rev.~A {\bf 93}, 063629 (2016).
\bibitem{Son} D.~T.~Son, Phys.~Rev.~B {\bf 75}, 235423 (2007).
\bibitem{Foster} M.~S.~Foster and I.~L.~Aleiner, Phys.~Rev.~B {\bf 77}, 195413 (2008).
\bibitem{Pastukhov} V.~Pastukhov, J.~Phys.~A:~Math.~Theor. {\bf 51} 195003 (2018).
\bibitem{Vasilev} A.~N.~Vasil'ev, Y.~M.~Pis'mak, Y.~R.~Khonkonen,  Theor.~Math.~Phys. {\bf 46}, 104 (1981); Theor.~Math.~Phys. {\bf 47}, 465 (1981).
%\bibitem{Yabunaka} S.~Yabunaka and B.~Delamotte, Phys.~Rev.~Lett. {\bf 119}, 191602 (2017).
\bibitem{Brezin} E.~Br\'{e}zin, C.~Itzykson, G.~Parisi, and J.~B.~Zuber, Commun.~Math.~Phys. {\bf 59}, 35 (1978).
\bibitem{Stanley} H.~E.~Stanley, Phys.~Rev. {\bf 176},  718 (1968).

\bibitem{Brezin_Wadia} E.~Br\'ezin and S.~R.~Wadia, {\it The Large N Expansion in Quantum Field Theory and Statistical Physics: From Spin Systems to 2-Dimensional Gravity}, (World Scientific, 1993).




\bibitem{Andersen} J.~O.~Andersen, arXiv:cond-mat/0608265.
\bibitem{Nogueira} F.~S.~Nogueira, 	arXiv:1009.1603.
\bibitem{Brauner} J.~O. Andersen and T.~Brauner, Phys.~Rev.~D {\bf 78}, 014030 (2008); J.~O. Andersen, arXiv:hep-ph/0501094. 
\bibitem{Chien} C.-C.~Chien, F.~Cooper, and E.~Timmermans, Phys.~Rev.~A {\bf 86}, 023634 (2012).

\bibitem{Liu} X.-J.~Liu, B.~Mulkerin, L.~He, and H.~Hu, Phys.~Rev.~A {\bf 91}, 043631 (2015).

\bibitem{Nogueira_Kleinert} F.~S.~Nogueira and H.~Kleinert, Phys.~Rev.~B {\bf 73}, 104515 (2006).


\bibitem{Hryhorchak} O.~Hryhorchak and V.~Pastukhov, EPL (Europhysics Letters) {\bf 118}, 56003 (2017).

\bibitem{Bronin} S.~Y.~Bronin, B.~V.~Zelener, A.~B.~Klyarfeld, V.~S.~Filinov,
Europhys.~Lett. {\bf 103},  60010 (2013).
\bibitem{Nguyen} T.~T.~Nguyen,  A.~J.~Herrmann, M.~Troyer, S.~Pilati, Phys.~Rev.~Lett. {\bf 112}, 170402 (2014).

\bibitem{Popov} V.~N.~Popov, {\it Functional Integrals and Collective Excitations} (Cambridge University Press, Cambridge, 1987).

\bibitem{Nepomnyashchy} A.~A.~Nepomnyashchy and Y.~A.~Nepomnyashchy, Pis'ma Zh.~Exp.~Teor.~Fiz. {\bf 21}, 3 (1975)
[Sov.~Phys.~JETP~Lett. {\bf 21}, 1 (1975)]; Y.~A.~Nepomnyashchy
and A.~A.~Nepomnyashchy, Zh.~Exp.~Teor.~Fiz. {\bf 75}, 976 (1978)
[Sov.~Phys.~JETP {\bf 48}, 493 (1978)].

\bibitem{Hugenholtz_Pines} N.~M.~Hugenholtz and D.~Pines, Phys.~Rev. {\bf 116}, 489 (1959).

\bibitem{Capogrosso-Sansone} B.~Capogrosso-Sansone, S.~Giorgini, S.~Pilati, L.~Pollet, N.~Prokof'ev, B.~Svistunov, and M.~Troyer, New~J.~Phys. {\bf 12} 043010 (2010).

\bibitem{Szepfalusy} P.~Sz\'epfalusy, I.~Kondor, Ann.~Phys. {\bf 82} 1 (1974).
\bibitem{Fliesser_etal} M.~Fliesser, J.~Reidl, P.~Sz\'epfalusy, and R.~Graham, Phys.~Rev.~A {\bf 64}, 013609 (2001).
\bibitem{Navez} P.~Navez, J.~Low~Temp.~Phys. {\bf 138}, 705 (2005); Physica~A {\bf 356}, 241 (2005).

\bibitem{Watabe} S.~Watabe and Y.~Ohashi, Phys.~Rev.~A {\bf 88}, 053633 (2013); Phys.~Rev.~A {\bf 90}, 013603 (2014).




\bibitem{Kita} T.~Kita,  J.~Phys.~Soc.~Jap. {\bf 74} 1891 (2005); J.~Phys.~Soc.~Jap. {\bf 75} 044603 (2006).
\bibitem{Yukalovs_1} V.~I.~Yukalov, E.~P.~Yukalova, Phys.~Rev.~A {\bf 76}, 013602 (2007).
\bibitem{Cooper_etal} F.~Cooper, B.~Mihaila, J.~F. Dawson, C.-C.~Chien, and E.~Timmermans, Phys.~Rev.~A {\bf 83}, 053622 (2011).
\bibitem{Zhang} Y.-H.~Zhang and D.~Li, Phys.~Rev.~A {\bf 88}, 053604 (2013).
\bibitem{Yukalovs_2} V.~I.~Yukalov, E.~P.~Yukalova, J.~Phys.~B: At. Mol. Opt. Phys. {\bf 47}, 095302 (2014).
\bibitem{Brewczyk_etal} M.~Brewczyk, M.~Gajda and K.~Rzazewski, J.~Phys.~B: At.~Mol.~Opt.~Phys. {\bf 40} R1 (2007).


\bibitem{Lopes_etal} R.~Lopes, C.~Eigen, N.~Navon, D.~Cl\'ement, R.~P.~Smith, and Z.~Hadzibabic, Phys.~Rev.~Lett. {\bf 119}, 190404 (2017).

\bibitem{Abe} R.~Abe and S.~Hikami, Prog.~Theor.~Phys. {\bf 49}, 442 (1973).

\bibitem{Campostrini_01} M.~Campostrini, M.~Hasenbusch, A.~Pelissetto, P.~Rossi, and E.~Vicari, Phys.~Rev.~B {\bf 63}, 214503 (2001).
\bibitem{Campostrini_06} M.~Campostrini, M.~Hasenbusch, A.~Pelissetto, and E.~Vicari, Phys.~Rev.~B {\bf 74}, 144506 (2006).

\bibitem{Vakarchuk_etal} I.~Vakarchuk, O.~Hryhorchak, V.~Pastukhov, R.~Prytula, Ukr.~J.~Phys. {\bf 61}, 29 (2016).

\bibitem{Hryhorchak_15} O.~I.~Hryhorchak, Condens.~Matter~Phys. {\bf 18}, 43001 (2015).

\bibitem{Boninsegni} M.~Boninsegni, N.~V.~Prokof'ev, and B.~V.~Svistunov,
Phys.~Rev.~E {\bf 74}, 036701 (2006).

\bibitem{Ville} J.~L.~Ville, R.~Saint-Jalm, \'E.~Le~Cerf, M.~Aidelsburger, S.~Nascimb\`ene, J.~Dalibard, J.~Beugnon, 	arXiv:1804.04037.

%\bibitem{Ferrell} R.~A.~Ferrell and D.~J.~Scalapino, Phys.~Lett.~A {\bf 41}, 371
%(1972); Phys.~Rev.~Lett. {\bf 29}, 413 (1972).

%\bibitem{Hore_Frankel} S.~R.~Hore and N.~E.~Frankel, Phys.~Rev.~B {\bf 12}, 2619 (1975).


\end{thebibliography}
\end{document}